\documentclass[twocolumn,showpacs,preprintnumbers,amsmath,amssymb]{revtex4}

\usepackage[T1]{fontenc}
\usepackage{graphicx}
\usepackage{dcolumn}
\usepackage{bm}
\usepackage[ansinew]{inputenc} 
\usepackage{subfigure} 
\usepackage{lscape} 

\usepackage{color,graphicx}
\usepackage{ifpdf}
\ifpdf 
  \DeclareGraphicsExtensions{.pdf,.png,.mps}
\else 
  \DeclareGraphicsExtensions{.eps,.bmp}
  \usepackage{pstricks}
\fi
\usepackage{epic,bez123,wrapfig}

\begin{document}

\title{Neutron stars with small radii - the role of $\Delta$ resonances}

\author{Torsten Schürhoff}
\affiliation{FIAS, Institute for Theoretical Physics, Johann
Wolfgang Goethe University, Frankfurt am Main, Germany}

\author{Stefan Schramm}
\affiliation{CSC, FIAS, ITP, Johann Wolfgang Goethe University,
Frankfurt am Main, Germany}

\author{Veronica Dexheimer}
\affiliation{Gettysburg College, Gettysburg, USA}

\date{\today}

\begin{abstract}

Recent neutron star observations suggest that the masses and radii
of neutron stars may be smaller than previously considered, which
would disfavor a purely nucleonic equation of state. In our model,
we use a the flavor SU(3) sigma model that includes
$\Delta$-resonances and hyperons in the equation of state. We find
that if the coupling of the $\Delta$ resonances to the vector
mesons is slightly smaller than that of the nucleons, we can
reproduce both the measured mass-radius relationship and the
extrapolated equation of state.

\end{abstract}

\maketitle

\section{Introduction}

The mass-radius relationship of neutron stars is of key importance
if one wants to understand the high-density low-temperature region
of the hadronic equation of state. Depending on this relationship,
certain models for the hadronic equation of state can either be
confirmed or ruled out.

A recent analysis of the masses and radii of neutron stars by Özel
et al. \cite{Oezel Main,Oezel Measurement 1, Oezel Measurement 2,
Oezel Measurement 3}
 seems to imply that they are smaller than previously assumed,
 with radii between 8 and
10 kilometers and masses between 1.6 and 1.9 solar masses. This
information is extracted from the thermonuclear bursts of x-ray
emitting neutron stars in binary systems. Using the measured
distance, the apparent surface area during the cooling phase of
the bursts and the flux in certain types of bursts which exceed
the Eddington limit, a constraint for the masses
and radii of the measured neutron stars is deduced. From this, a
 backward-extrapolation of error bars for equations of state \cite{Oezel
Reconstruction,Harada Reconstruction,Reconstruction Lindblom,
Reconstruction Prakash} that are in accordance with the measured
values is performed.
 A comparison with various models of
equations of state suggests that a purely nucleonic equation of
state is too stiff, i.e. produces too high masses at too large
radii. It is therefore possible that more exotic particles are
realized within the interior of neutron stars \cite{FWeber1},
which soften the equation of state.

Several attempts have already been made to understand the measured
values. Fattoyev et. al \cite{Fattoyev} use RMF models with a
nucleonic EoS. While they can fit the EoS, they are not able to
fully reach the small masses and radii reported by Özel et al. .
Ciarcelluti et al. \cite{DSCore} explore the possibility of
neutron stars having a dark matter core. Drago et al. \cite{Drago}
propose that this could be an experimental signal for the
existence of two types of stars, ordinary neutron stars and
``ultra-compact`` neutron stars like quark stars. Steiner et al.
\cite{Steiner}
 performed a reanalysis of the experimental data with slightly different assumptions.
They relax the assumption of the stellar radius at maximum temperature
being equal to the photospheric radius, increasing the allowed
radius interval. As a result, they reproduce mass-radius
relationships which are in agreement with a purely nucleonic
equation of state.

In our approach we want to explore the possibility of additional
particle species being present, namely $\Delta$ resonances and
hyperons, and investigate the effect on the equation of state of
neutron stars.

\section{Description of the model}

A hadronic SU(3) sigma-omega model is used to describe the
properties of neutron star matter. The included baryonic degrees
of freedom are the baryon octet (n,p, and hyperons $\Lambda,
\Sigma^{+,0,-}, \Xi^{0,-}$), the leptons (e, $\mu$), and the
resonances from the spin 3/2 decuplet ( $\Delta^{++,+,0,-}$,
$\Sigma^{*+,0,-}$, $\Xi^{*0,-}$, and $\Omega^{-}$). The mesons,
that mediate the interactions between the baryons, are the
vector-isoscalar $\omega$ and $\phi$, the vector-isovector $\rho$
and the scalar-isoscalar $\sigma$ and $\zeta$ (strange
quark-antiquark state).

The Lagrangian density reads

\begin{eqnarray}
&L = L_{Kin}+L_{Int}+L_{Self}+L_{SB}.&
\end{eqnarray}

$L_{Kin}$ is the kinetic energy term for the hadrons and leptons.
In addition there is an interaction term between the baryons and
the scalar and vector mesons

\begin{eqnarray}
&L_{Int}=-\sum_i
\bar{\psi_i}[\gamma_0(g_{i\omega}\omega+g_{i\phi}\phi+g_{i\rho}\tau_3\rho)+M_i^*]\psi_i\nonumber,&\\&
\end{eqnarray}
with the effective mass $M_i^*$ given by
\begin{eqnarray}
&M_i^* = g_{i\sigma} \sigma + g_{i\zeta} \zeta + \delta m_i  ,&
\end{eqnarray}
with a small bare mass term $\delta m_i$. The coupling strengths
of the baryons to the scalar fields are connected via SU(3)
symmetry relations and the different SU(3) invariant coupling
strengths are fitted to reproduce the baryon masses in vacuum (see
Ref. \cite{Veronica Hybrid stars} for the values of the
couplings).

The self-interaction terms for the scalar and vector mesons read,

\begin{eqnarray}
&L_{Self}=-\frac{1}{2}(m_\omega^2\omega^2+m_\rho^2\rho^2+m_\phi^2\phi^2)\nonumber&\\&
+g_4\left(\omega^4+\frac{\phi^4}{4}+3\omega^2\phi^2+\frac{4\omega^3\phi}{\sqrt{2}}+
\frac{2\omega\phi^3}{\sqrt{2}}\right)\nonumber&\\&+k_0(\sigma^2+\zeta^2)
+k_1(\sigma^2+\zeta^2)^2&\nonumber\\&+k_2\left(\frac{\sigma^4}{2}+\zeta^4\right)
+k_3\sigma^2\zeta+k_4\
\ln{\frac{\sigma^2\zeta}{\sigma_0^2\zeta_0}},&
\end{eqnarray}

\noindent and the explicit chiral symmetry breaking term is given
by

\begin{eqnarray}
&L_{SB}= m_\pi^2 f_\pi\sigma+\left(\sqrt{2}m_k^
2f_k-\frac{1}{\sqrt{2}}m_\pi^ 2 f_\pi\right)\zeta.
\end{eqnarray}

In the case of the baryonic octet we use an f-type coupling between the baryons and
vector mesons, which yields coupling strengths as given by quark counting rules, i.e.
$g_{i\omega} = (n^i_q-n^i_{\bar{q}}) g_{8}^V$\,,
$g_{i\phi}   = -(n^i_s-n^i_{\bar{s}}) \sqrt{2} g_{8}^V$\,,
where 
$g_8^V$ denotes the vector coupling of the baryon
octet and
$n^i$ the number of constituent quarks of species $i$ in a given hadron.
A more detailed description of the model can be found in
\cite{Papazouglou}. The model has been tested extensively both
for nuclear matter \cite{Zeeb Phase structure} and neutron stars
 \cite{Veronica Hybrid stars,Veronica Parity Doublet}.

The parameters are fitted to reproduce the correct vacuum masses
of the baryons and mesons,
 nuclear saturation properties at
$\rho_{0}= 0.15$ particles per $\rm{fm^{3}}$ (the correct binding
energy per nucleon, B/A = - 16 MeV and a compressibility of K =
297.32 MeV) and the asymmetry energy ($E_{sym}=32.5 \rm{MeV}$).

As the values of the vector coupling of the decuplet are not
constrained by the properties of saturated nuclear matter as in
the case of the baryon octet, in the following we allow for
moderate deviations from the overall vector coupling strength for
the baryon decuplet compared to the octet. In order to study the
effects of such a deviation in a most direct way we introduce a
single parameter $r_v$, which in the case of the $\Delta$
resonances is defined as

\begin{equation}
r_v =  \frac{g_{\Delta\omega}}{g_{N\omega}}
\end{equation}

with the same rescaling of the other states of the decuplet.

Using this approach we solve the equations of motion by differentiating the
grand-canonical potential $\frac{\Omega}{V}=-p = \epsilon -
\mu\varrho_{B}$ of the model for $T=0$ with respect
 to the fields (in mean-field approximation).
Then we determine the equation of state and use it to solve the
Tolman-Oppenheimer-Volkoff equations for spherical static stars. A
standard BPS crust EoS \cite{crust} is added to the outer layers
of the star.

\section{Results}

$\Delta$ resonances have a repulsive vector potential which works
to counteract gravity in a compact star. If the interaction
strength of the $\Delta$ resonances with respect to the nucleons
is lowered, i.e. $r_{v}<1$, then the equation of state becomes
softer and hence the maximum supported mass of the star decreases.

\begin{figure}
\centering \vspace{1.0cm}
\includegraphics[width=8.cm]{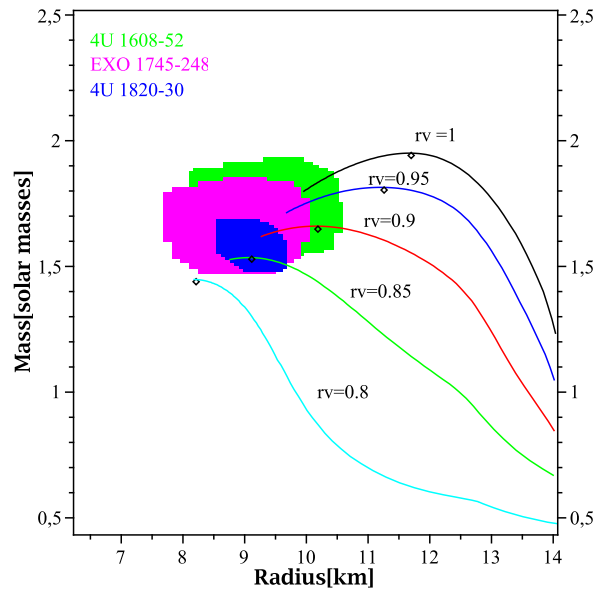}
\vspace{-0.4cm} \caption{\label{fig1} Mass-Radius-relationship of
Neutron stars for various couplings of the $\Delta$ resonances,
starting from $r_{v}=1$ (upper line) to 0.8 (lowest line). Also
included are the 1-$\sigma$ errorbars for measured neutron stars
from \cite{Oezel Main}. The black diamond on each curve represents
the maximum stable configuration of the neutron star.}
\end{figure}

Figure 1 shows the impact of various coupling strengths $r_{v}$ of
the $\Delta$ resonances on the overall mass-radius relation of the
neutron star. The upper black curve represents $r_v=1$, and in
each subsequent curve the coupling strength is lowered by 0.05,
arriving at 0.8 in the lowest curve. If the coupling strength of
the $\Delta$ resonances is reduced, the maximum mass of the star
likewise decreases. We found that for values of 0.7 or lower, the
equation of state has to be considered with care and may start to
yield unphysical results such as the appearance of $\Delta$s in
the nuclear ground state.

However, for values of $r_v$ around 0.9, which corresponds to the
physical case where the coupling strength of the $\Delta$
resonances is only slightly lowered compared to normal nucleons,
the mass-radius-relation is in good agreement with the values
given by Özel et al. .

Figure 2 shows the equation of state for various cases considered.
For comparison reasons, we use the same units as Özel et al.
\cite{Oezel Main} and include their error bars (black
perpendicular lines). The stiffest line represents a purely
nucleonic equation of state, n, p and e being present. In all
subsequent cases $\Delta$ resonances are included. We look at two
cases, $r_v=1$ and $r_v=0.9$ (both dashed). The dotted lines below
each of the curves represents the effects of muons and hyperons
being added. The inclusion of additional particle species
generally softens the equation of state. The major effect comes
from the delta resonances, while the hyperons and muons only make
minor changes.

\begin{figure}
\centering \vspace{1.0cm}
\includegraphics[width=8.cm]{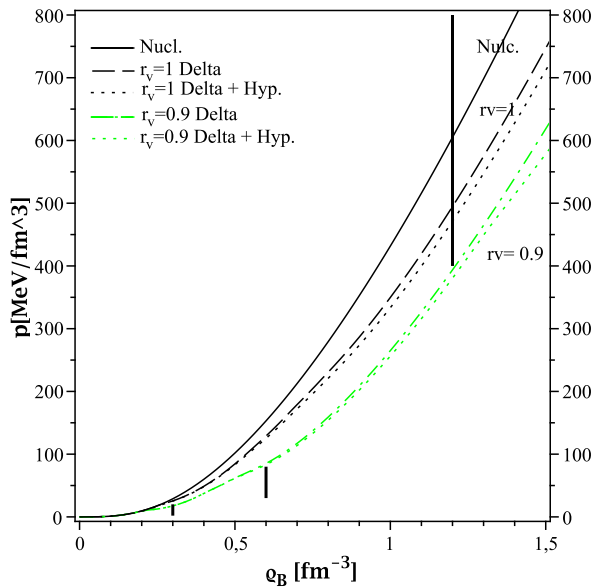}
\vspace{-0.4cm} \caption{\label{fig2} Considered equations of
state(from stiffest to softest): purely nucleonic (solid),
including $\Delta$ resonances at $r_v=1$ (dashed), $\Delta$
resonances and hyperons at $r_v=1$(dotted), $\Delta$ resonances at
$r_v=0.9$ (dashed), $\Delta$ resonances and hyperons at $r_v=0.9$
(dotted). The additional inclusion of hyperons only creates a
minor effect.}
\end{figure}

\begin{figure}
\centering \vspace{1.0cm}
\includegraphics[width=8.cm]{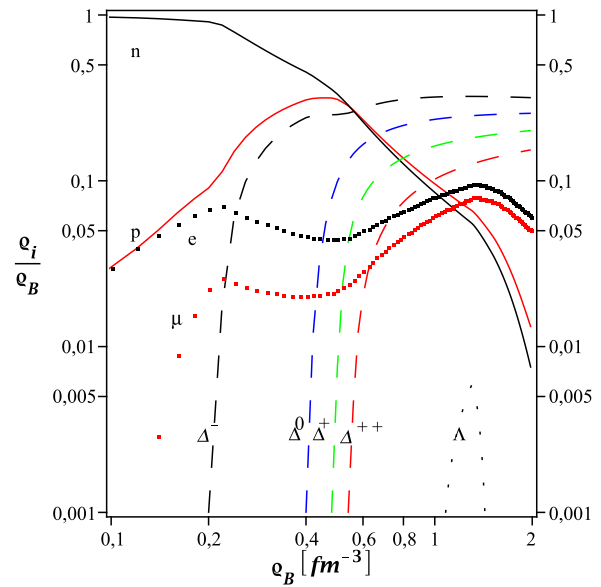}
\vspace{-0.4cm} \caption{\label{fig3}Particle abundancy as a
function of the baryonic density in $1/fm^3$ for the case with
hyperons and Delta resonances present, rv =0.9 .}
\end{figure}

Figure 3 shows the particle abundances for the best-fit case of
$r_v = 0.9$, with nucleons, leptons, $\Delta$ resonances and
hyperons present. At densities between $0.2$ and $0.5$ particles
per $\rm{fm^3}$, where $\varrho_{0}=0.16 fm^(-3)$ represents
normal nuclear matter density, the $\Delta$ resonances start to
appear, first the $\Delta^-$ then the others. At very high
densities beyond six times $\varrho_{0}$, the hyperons start to
appear. In principle all the listed particles are present in the
model and can appear, including the excited spin 3/2 states.
However, as one can see in Figure 3, only the $\Lambda$ starts to
appear in a small fraction in the relevant regime.

It is noteworthy that negatively charged particles like the muon
and the $\Delta^{-}$ at a certain point take up the role of the
electrons in charge conservation and thus reduce the electron
abundance. At a certain point, the $\Delta$ particles, due
to their reduced repulsive potential, become the dominant particle
species.

\begin{figure}
\centering \vspace{1.0cm}
\includegraphics[width=8.cm]{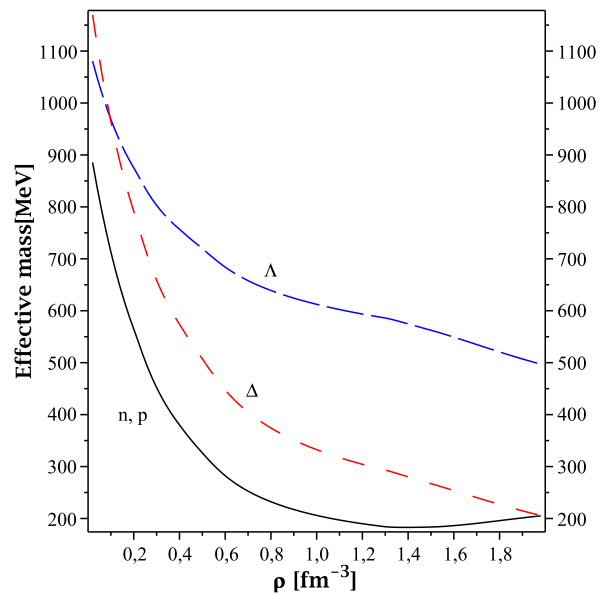}
\vspace{-0.4cm} \caption{\label{fig4}Effective mass of the
particles as a function of the baryonic density.}
\end{figure}

Figure 4 shows the effective mass of the nucleons, $\Delta$ and
$\Lambda$ particles within the chiral model as a function of the
density.

\section{Discussion}

Recent experimental data \cite{Oezel Main} seem to favor neutron
star masses and radii that are smaller than previously considered.
An equation of state consisting only of nucleons and electrons is
not able to reproduce the measured masses and radii.

Note, however, that in a different analysis, Steiner et al.
\cite{Steiner} relax the assumption of the stellar radius at
maximum temperature being equal to the photospheric radius,
increasing the allowed radius interval. As a result, they
reproduce mass-radius relationships which are in agreement with a
purely nucleonic equation of state. If the analysis of Oezel et
al. is correct, and neutron stars have smaller masses and radii
than previously considered, then this may hint to more exotic
particle species being present in the equation of state. We have
shown that an equation of state with $\Delta$ resonances in such a
case would be in accordance with the experimental data. While our
approach includes hyperonic degrees of freedom in a natural way,
the results are dominated by the impact of the $\Delta$
resonances, whereas hyperons only have a minor effect on the
result.

If, on the other hand, Steiner et al. should be correct with their
analysis, this would point to a more conservative mass radius
relation and equation of state, which can be described by the
presence of nucleons and electrons alone. A greater amount of
analyzed experimental data may be necessary to resolve this issue.

\section{Conclusions}

We have applied a non-linear realization of an extended
sigma-omega model, including $\Delta$ resonances and hyperons in
the equation of state of a neutron star calculation. A parameter
scan for the coupling strength of the $\Delta$ resonances has been
performed. By slightly reducing the strength of the vector
coupling by about ten percent compared to the values for the
baryonic octet we obtain good agreement with the results from Ref.
\cite{Oezel Main}, both in the mass-radius relation and equation
of state. It is noteworthy that this agreement can  be achieved by
the not too-exotic inclusion of $\Delta$ resonances in the
equation of state without considering quark cores or quark stars.

\section{Acknowledgements}
We thank Rodrigo Negreiros for very fruitful discussions. We
gratefully acknowledge the access to the computing resources of
the Center for Scientific Computing (CSC), Frankfurt University,
for our calculations.


\begin{thebibliography}{99}

\bibitem{Oezel Main} Feryal Özel, Gordon Baym, Tolga Güver,
   \emph{Astrophysical measurement of the equation of state of neutron star
   matter}, astro-ph.HE/1002.3153v1 (2010)

\bibitem{Oezel Measurement 1} Tolga Güver, Patricia Wroblewski, Larry Camarota, Feryal Özel,
   \emph{The mass and radius of the neutron star in U4 1820-30},
    astro-ph.HE/1002.3825v1 (2010)

\bibitem{Oezel Measurement 2} Feryal Özel, Tolga Güver, Dimitros Psaltis,
\emph{The mass and radius of the neutron star in EXO 1745-248},
 ApJ 693, 1775 (2008)

\bibitem{Oezel Measurement 3}
Tolga Güver, Feryal Özel, Antonio Cabrera-Lavers, Patricia
Wroblewski, \emph{The Distance, Mass, and Radius of the Neutron
Star in 4U 1608-52},
 ApJ 712, 964 (2010)

\bibitem{Oezel Reconstruction} Feryal Özel, Dimitros Psaltis,
 \emph{Reconstructing the neutron star
equation of state from astrophysical measurements},
 Phys.Rev.D 80, 103003 (2009)

\bibitem{Harada Reconstruction} Tomohiro Harada, \emph{Reconstructing the equation of state for cold nuclear matter
from the relationship of any two properties of neutron stars},
Phys.Rev.C 64, 048801 (2001)

\bibitem{Reconstruction Lindblom} Lee Lindblom, \emph{Determing the nuclear equation of state from neutron-star
 masses and radii}, Astrophys.J. 398,569-573, (1992)

\bibitem{Reconstruction Prakash} James M. Lattimer, Maddapa Prakash,
\emph{Neutron Star Observations: Prognosis for Equation of State
Constraints}, 
Physics Reports Volume 442, 106-165 (2007)

\bibitem{FWeber1} Fridolin Weber, \emph{Strangeness in neutron
stars}, 
J.Phys.G.:Nucl.Part.Phys. 27-465 (2001)

\bibitem{Fattoyev} F.J. Fattoyev, J. Piekarewicz, \emph{Relativistic models of the neutron star matter equation of
state}, nucl-th/1003.1298v1 (2010)

\bibitem{DSCore} Paolo Ciarcelluti, Frederik Sandin, \emph{Have neutron stars a dark matter
core?}, astro-ph.HE/1005.0857v1 (2010)

\bibitem{Drago} Allessandro Drago, Andrea Lavagno, \emph{Masses and radii of neutron and quark
stars}, astro-ph.SR/1004.0325v1 (2010)

\bibitem{Steiner} Andrew W. Steiner, J. Lattimer, Edward F. Brown
\emph{The equation of state from observed masses and radii of
neutron stars}, astro-ph.HE/1005.0811v1 (2010)

\bibitem{Papazouglou} P. Papazouglou, D. Zschiesche, S. Schramm,
J.Schaffner-Bielich, H. Stöcker, W. Greiner, Phys. Rev.C. 59,411
(1999)

\bibitem{Zeeb Phase structure} D. Zschiesche, G.Zeeb, S. Schramm,
\emph{Phase structure in a hadronic chiral model},
J.Phys.G.: Nucl.Part.Phys. 34 1665 (2007)


\bibitem{Veronica Hybrid stars} V. Dexheimer, S. Schramm,
\emph{A novel approach to model hybrid stars}
Phys.Rev.C 81,045201 (2010)


\bibitem{Veronica Parity Doublet} V. Dexheimer, S. Schramm, D. Zschiesche,
\emph{Nuclear matter and neutron stars in a parity doublet model},
Phys.Rev.C 77, 025803 (2008)

\bibitem{crust}
 G.~Baym, C.~Pethick and P.~Sutherland,
 \emph{The Ground State Of Matter At High Densities: Equation Of State And Stellar
 Models}
 Astrophys.\ J.\  170, 299 (1971).

\bibitem{DeltaResonancesGoodwin} N.H. Goodwin,\emph{The potential energy of an infinite system of nucleons and
delta resonances}, J.Phys.G:Nucl. Phys.6  815-839 (1980)




\end{thebibliography}
\end{document}